# Globular Cluster Systems as Probes of Elliptical Galaxy Formation[1]

Stephen E. Zepf[2]

Department of Astronomy, University of California

Berkeley, CA 94720

## ABSTRACT

Globular cluster systems provide valuable fossil records of the formation history of their parent galaxies. This review specifically concentrates on using color distributions of the globular cluster systems of elliptical galaxies to distinguish between competing models for the formation of these galaxies. The observational requirements for testing various formation models are discussed, and the results of past and current studies are reviewed. The conclusions of these studies are that most color distributions appear to have two or more peaks, indicating an episodic formation history. This result is consistent with previous predictions of merger models for the formation of elliptical galaxies. Possibilities for improving these results in the future are also presented.

## 1. Introduction

The study of globular cluster systems (GCSs) has long been at least partially motivated by the idea that GCSs can be used as fossil records of the formation history of their host galaxies (e.g. Harris & Racine 1979, Harris 1991). Ashman & Zepf (1992) placed this idea in the context of merger models for elliptical galaxy formation. They suggested that galaxy interactions and mergers are favorable environments for globular cluster formation (see also Schweizer 1987). This proposal now appears to be confirmed by HST imaging studies of galaxy mergers, which find large numbers of objects with the characteristics expected of young globular clusters (see review by Whitmore in these proceedings). Moreover, in the few cases for which spectra have been obtained, strong Balmer absorption lines are observed, supporting the identification of these objects as young globular clusters (Schweizer & Seitzer 1993, Zepf et al. 1995a).

Ashman & Zepf (1992) noted that if globular clusters form in mergers of gas-rich galaxies, then elliptical galaxies formed by mergers will have GCSs composed of two or more distinct populations of globular clusters. The halo clusters of the progenitor spirals form one population which is metal-poor and spatially extended. A second population is composed of globular clusters formed during the merger. These clusters will be younger, and also more metal-rich than the halo

---

population, since they form out of material which has been enriched in the disks of the progenitor spirals and in the merger itself. Because most mergers are likely to occur at moderate or high redshifts, metallicity differences play a much larger role than age differences in determining the colors and metallic absorption-line strengths for typical elliptical galaxy GCSs (Zepf & Ashman 1993). Thus, the clusters associated with the original spirals are expected to be bluer and have weaker metal-lines than the clusters formed during the merger in which the elliptical galaxy formed.

This prediction of two or more observably distinct populations in the GCS of an elliptical galaxy formed by a merger is in direct contrast to the single population predicted by monolithic collapse models. A cluster population formed in a monolithic collpase is expected to be roughly coeval and have a smooth, single-peaked metallicity distribution (e.g. Arimoto & Yoshii 1987). Observations of the distribution of colors or line-strengths for GCSs of elliptical galaxies provide a way to distinguish between monolithic collapse and merger models of the formation of elliptical galaxies. Although the formation of galaxies is probably much more complex than the scenarios described above, the general trends seem secure. More than one peak in the color distribution indicates episodic formation as expected in a merger picture, and single-peaked distributions are suggestive of monolithic collapse models.

Advances in observational capabilities have begun to allow a detailed examination of the color distributions of the GCSs of a number of elliptical galaxies. Thus, the time is ripe to determine the role of mergers in the formation of elliptical galaxies by studying their GCSs. The plan of this review is to first describe the data required to make this test, then to present past and current observational studies of the distribution of colors and line-strengths for GCSs, and to conclude by discussing the prospects for improving on these in the future.

## 2. Observational Requirements

A useful place to start is a consideration of the requirements of GCSs observations if they are to distinguish between merger and monolithic collapse models for the formation of elliptical galaxies. Perhaps the most basic issue is the detection of the globular clusters. Globular clusters are faint at even the closest distances for which samples of elliptical galaxies can be studied. For example, at the distance of Virgo, a typical Galactic globular cluster is $m_V \simeq 24$ and the brightest few percent are $m_V \simeq 21$. Although they are faint, they are also numerous, so that observations which reach near the peak can expect to detect surface densities on the order of tens of objects per $\Box'$. Because the faintness of the objects has so far precluded the possibility of spectroscopically confirming large samples, globular clusters systems are typically defined by this overabundance of objects with the colors and magnitudes expected of globular clusters.

A second issue is that globular clusters are compact in appearance, with half-light radii of about $0.1''$ at the Virgo distance. Thus, detection benefits greatly from high resolution imaging.

High quality imaging is also valuable for distinguishing between globular clusters and background galaxies, because nearly all background galaxies are significantly more extended in appearance than globular clusters at the distance of Virgo. A third consideration is that the spatial extent of GCSs is usually at least as great as that of the underlying galaxy. Therfore, wide field imaging is very valuable for building sample size to allow for statistically significant results.

The use of the color distributions to distinguish between competing elliptical galaxy formation models requires more than detection alone. The requirements for the precision of the colors are set by the expected difference in metallicity between the populations in the merger picture, the intrinsic width of the metallicity distribution of the individual populations, and the age difference between the two populations. The assumptions of two roughly coeval populations with no instrinsic metallicity dispersion and a separation of $\Delta[\mathrm{Fe/H}] = 1.0$ provide a useful starting point. A simple expectation is that two peaks can be detected only if they are separated by at least twice the standard deviation of the individual peaks (Everitt & Hand 1981). This would require determining metallicities to a precision of $[\mathrm{Fe/H}] \leq 0.50$.

There are many factors which mitigate this optimstic estimate of the required photometric precision (Ashman, Bird, & Zepf 1994). For example, Ashman et al. (1994) showed that with sample sizes of 100-200, which are typical for currently available data, a metallicity precision of a factor of 2.5-3.0 less than the expected separation is required for a reliable detection or rejection of bimodality using the best available algorithims. For typical conversions of $\Delta[Fe/H] = 1.0$ to color indices (e.g. Couture, Harris, & Allwright 1990, Worthey 1994), this precision corresponds to $\sigma(V - I) \leq 0.07$, or $\sigma(B - I) \leq 0.13$. As the photometric precision improves, one limit for the detection of individual populations is internal metallicity dispersion of the populations themselves. In the case of the halo globular clusters of the Galaxy, the dispersion is about $\sigma[Fe/H] = 0.33$ (Armandroff & Zinn 1988), or $\sigma(B - I) = 0.12$.

Age differences between the two populations are also expected at some level. Their effect will be to make bimodality harder to detect, since the higher metallicity clusters formed during mergers will also be younger. Fortunately, as discussed in Section 1, the effect is not likely to be significant in most cases, because most mergers must occur at fairly high redshift if elliptical galaxies are to be made this way. To take a specific example, a solar metallicity population is expected to be about 0.1 mag bluer in $(B - I)$ after 10 Gyr than after 15 Gyr (Bruzual & Charlot 1996). In order for the age difference to match the effect in $(B - I)$ of a metallicity difference of 1.0 dex, the younger age must be roughly 3 Gyr if the older age is 15 Gyr. Although age-dating stellar populations is still a fairly uncertain business (Charlot, Worthey, & Bressan 1996), it is likely that the predictions of *relative* ages are more robust.

The observational requirements to obtain adequate data for testing whether elliptical galaxies formed by merging through the color distribution of their GCSs can be roughly summarized as $\lesssim 10\%$ photometry in two bandpasses, down to a limiting magnitude $m_V \sim 24$, and over an area of about 10 $\Box'$. Within this rough outline, area and depth can be traded off to some extent. Very approximately, doubling the radius gives an equivalent increase in number of clusters as doubling the sensitivity. This approximation breaks down badly at large radii when background contamination becomes significant, for very bright magnitudes at which there is no cluster population, and at faint magnitudes when the turover in the LF begins to be reached. Of course, there are advantages to deeper data, with the improved precision for brighter objects, and for data over a wider field, with a better handle on radial variations in GCS properties.

### 3. Past

The history of tests of the formation history of elliptical galaxies through studies of the color distribution of their GCSs is short. This is not surprising, given the observational requirement discussed above. Moreover, strong observational evidence for the formaton of globular cluster in mergers has only become available within the past few years. The first statistical test for bimodality in the color distributions of elliptical galaxy GCSs was presented by Zepf & Ashman (1993). They utilized the best available photometry at the time, which was 72 $B - I$ colors for NGC 4472 globular clusters (Couture et al. 1991), and 60 $C - T_1$ colors for NGC 5128 globular clusters (Harris et al. 1991). Based on a mixture modeling analysis, Zepf & Ashman found a bimodal distribution was preferred to a unimodal one at a 98.5% confidence level for the NGC 4472 GCS and a 95% confidence level for the NGC 5128 GCS.

This result was intriguing, as it suggested that these elliptical galaxies were likely to have formed through merging. At the same time, clearly better data, and in particular larger samples of globular clusters, were required to decide the issue without question. Fortunately, these earlier data were all obtained with small format CCDs, and larger format CCDs would bring significant improvements. Among the next generation of studies were those of Doug Geisler and collaborators, who analyzed the color distrbution of the GCSs of NGC 1399 (Ostrov, Forte, & Geisler 1993), and M87 (Lee & Geisler 1993). In both cases, a statistical analysis indicated that the distributions had two or more peaks. Similarly, Ajhar, Blakeslee, & Tonry (1994) surveyed a number of early-type galaxies, and obtained $V - I$ colors for 100 or more clusters around five Virgo ellipticals. A statistical analysis of these data showed that two of the GCSs are bimodal with high confidence, one is likely to be bimodal, and two appear to be unimodal (Zepf, Ashman, & Geisler 1995b).

Further evidence that bimodal color distributions are common for the GCSs of elliptical galaxies came from the studies of NGC 3311 by Secker et al. (1995) and of NGC 3923 by Zepf et al. (1995b). In addition to color distributions which were bimodal, the GCSs of both of these galaxies have fewer blue clusters than that the M87 GCS. This is particularly interesting in the case of NGC 3923, which is fainter than M87, and thus indicates that there is not a monotonic relationship between elliptical galaxy luminosity and GCS color or inferred metallicity (Zepf et al. 1995b). They also showed that the the detection of two or more peaks in the color distribution was robust to the non-linearities in the color-metallicity relation predicted by theory.

## 4. Present

The studies discussed in the previous sections demonstrated that color distributions of elliptical galaxy GCSs are typically bimodal rather than unimodal. Because bimodality was predicted to result from mergers and is difficult to understand in a monolithic collapse picture, these data alone provide a strong argument in favor of a merger origin for elliptical galaxies (Zepf et al. 1995b). At the same time, there is room for improvement on these earlier data. In particular, significantly increasing the number of clusters with good photometry allows a more detailed study of the individual peaks in detail. Deeper data also probes intrinsically fainter clusters to insure that they follow trends established for the brighter end of the population. Moreover, there is some background contamination, which, although at a low level, is still a source of noise.

Dramatic improvements in all of these areas were achieved with the analysis of WFPC2 images of the M87 GCS by Whitmore et al. (1995). They showed with unprecendented clarity that the color distribution of the M87 GCS divides into two distinct populations (see their Figure 4). One reason for the clear visual indication of two distinct peaks is the large number of clusters detected ($\sim 1000$) with high photometric precision. This wealth of clusters is due in large part to the depth which can be reached with HST imaging of compact sources. An additonal factor is the richness of the M87 GCS system itself, which makes for a high surface denisty of objects. The WFPC2 images also marginally resolve globular clusters at the distance of Virgo, providing a clean distinction between the clusters and background galaxies, which are much more extended, and even foreground stars, which are point sources. It is also interesting to note that a similar bimodal distribution is found for the M87 GCS at larger radii by Elson and collaborators.

To give a comparison of the HST data of Whitmore et al. (1995) for the M87 GCS to earlier work on other typical galaxies, their M87 GCS color histogram and the NGC 3923 GCS color distribution of Zepf et al. (1995b) is shown in Figure 1. As this figure demonstrates, each GCS clearly has a bimodal color distribution, indicative of an episodic formation history as expected in a merger scenario. However, the M87 data present a clearer picture, primarily because of the greater numbers and also because of the better control of contaminating objects.

## 5. Future

With the existence of bimodal color distributions firmly established, a natural question to ask is what is the next step. One obvious possibility is to conduct a systematic survey of elliptical galaxy GCSs to determine the frequency with which bimodality is detected and the relative number of objects in each peak, as well as the separation between peaks. Even in a simple merger model, bimodality is unlikely to be detected in all cases, as some combinations of metallicity and age differences will lead to colors which are too similar to separate. One step in this direction is thw work of Forbes et al. (1996) who used short exposure WFPC2 $V$ and $I$ images to study the GCSs of a number of nearby ellipticals. Because of the short exposures, the number of clusters around each elliptical galaxy with sufficiently precise photometry in $(V - I)$ is marginal for a statistical analysis of the color distributions. However, these data provide a useful baseline for comparison among galaxies and point the way for future surveys.

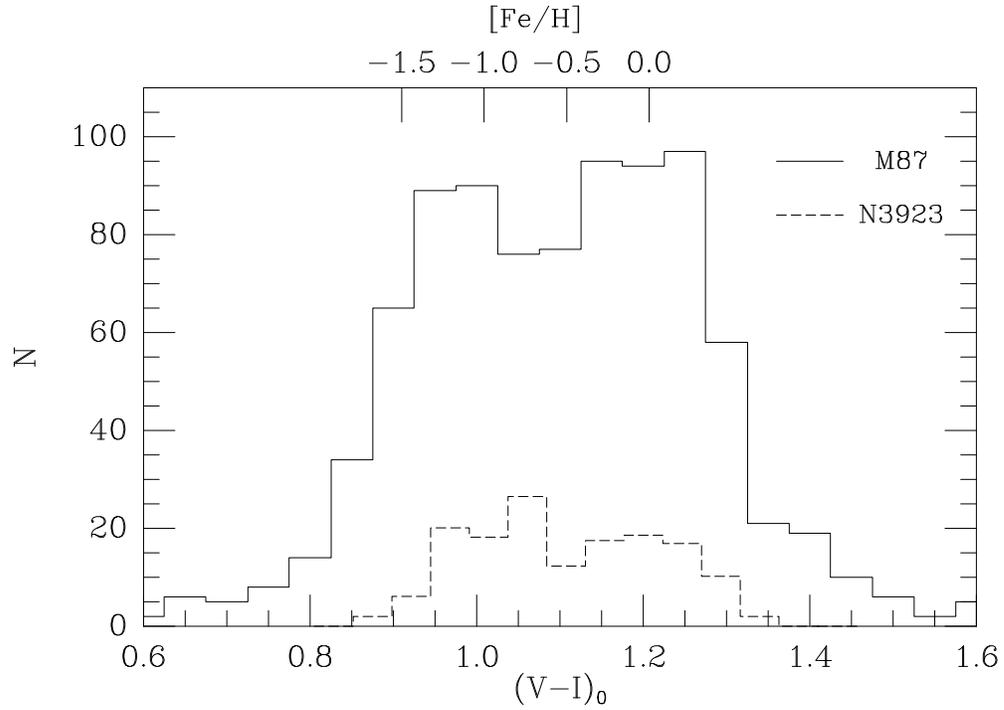

Fig. 1.— A histogram of the color distribution of the M87 GCS from Whitmore et al. (1995) and the NGC 3923 GCS from Zepf et al. (1995b), where the NGC 3923 color has been transformed to $(V-I)_0$ for comparison to the M87 data. Both distributions are bimodal. The overall color of the NGC 3923 GCS is somewhat redder, even though it is based on clusters at larger radii, and both of these GCSs have a color gradient which gets bluer with increasing radius.

Such surveys need not be performed with HST. Wide-field CCD images with good seeing on a large telescope can provide more than sufficient data for an analysis of the color distribution. The best current example is the recent study of the NGC 4472 GCS by Geisler, Lee, & Kim (1996). In roughly 3 hours of imaging through the Washington $C$ and $T_1$ fitlers with the 4m at KPNO, they detected well over 1000 clusters with a typical precision in terms of [Fe/H] of 0.15 dex. Geisler et al. (1996) strongly confirm the previous detection of bimodality in the NGC 4472 GCS, and are able to constrain the radial profiles of the two populations separately. Although background contamination remains somewhat of a concern for ground-based imaging surveys, it can be checked by HST imaging or ground-based spectroscopy of selected subset of the whole survey.

Spectroscopy opens up new possibilities for using GCSs to understand the formation history of elliptical galaxies. Firstly, redshifts establish on an individual basis a candidate cluster's association with its parent galaxy. The observed metal-line strengths also provide an independent way to search for distinct populations in the GCSs. Moreover, color and/or line strength information can be combined with the observed velocities to examine the kinematics of the different populations in a GCS. With higher quality spectra, one can also attempt to determine the [Mg/Fe] ratio of the globular clusters in much the same way as is done for the integrated spectra of elliptical galaxies (Worthey, Faber, & Gonzalez 1992, Davies, Sadler, & Peletier 1993). The [Mg/Fe] ratio is useful for constraining formation models because Mg is produced solely by SNII, whereas Fe is produced in SNI as well. Thus [Mg/Fe] ratios reflect the stellar initial mass function and the timescale of formation.

The recent availability of efficient multi-slit spectrographs and the advent of larger telescopes are making it feasible to obtain spectroscopy of large numbers of objects in the GCSs of elliptical galaxies. In addition to the motivations described above, spectra of globular clusters are of interest for probing the mass of elliptical galaxies at large radii (e.g. Zepf 1995, Grillmair et al. 1994, Brodie & Huchra 1991, Mould et al. 1990). The potential for simultaneously addressing the mass distribtion and the formation of history of elliptical galaxies with one set of observations makes spectroscopy of elliptical galaxy GCSs a particularly bright prospect for the future.

I thank my collaborators on the various projects described above, particularly Keith Ashman. I acknowledge Brad Whitmore for providing data in tabular form, and Becky Elson, Duncan Forbes, and Doug Geisler for sharing results prior to publication. My research is supported by NASA through grant number HF-1055.01-93A awarded by the Space Telescope Science Institute, which is operated by the Association of Universities for Research in Astronomy, Inc., for NASA under contract NAS5-26555. I also gratefully acknowledge partial travel support from the organizers of this meeting, and their gracious hospititality.